\documentclass[twocolumn,prb]{revtex4}
\usepackage[utf8]{inputenc}
\usepackage{amsmath,amsfonts,amssymb,graphicx,multirow}

\begin{document}
\title{$\mathbb{Z}_3$ Topological Order in Face Centered Cubic Quantum Plaquette Model}
\author{Trithep Devakul}
\affiliation{Department of Physics, Princeton University, Princeton 08540}
\date{\today}

\begin{abstract}
    We examine the topological order in the resonating singlet valence plaquette (RSVP) phase of the hard-core quantum plaquette model (QPM) on the face centered cubic (FCC) lattice.
    To do this, we construct a Rohksar-Kivelson type Hamiltonian of local plaquette resonances.
    This model is shown to exhibit a $\mathbb{Z}_3$ topological order, which we show by identifying a $\mathbb{Z}_3$ topological constant (which leads to a $3^3$-fold topological ground state degeneracy on the $3$-torus) and topological point-like charge and loop-like magnetic excitations which obey $\mathbb{Z}_3$ statistics.
    We also consider an exactly solvable generalization of this model, which makes the geometrical origin of the $\mathbb{Z}_3$ order explicitly clear.
    For other models and lattices, such generalizations produce a wide variety of topological phases, some of which are novel fracton phases.
    
\end{abstract}

\maketitle

\section{Introduction}

Quantum spin liquid phases~\cite{qslreview} are characterized by exotic behavior including emergent gauge fields and quasiparticle excitations which exhibit properties such as symmetry fractionalization and spin-charge separation.
Such phases are prime examples of topological order~\cite{wen}, which can be characterized by their topological ground state degeneracy~\cite{haldane,wen2}, non-trivial quasiparticle statistics~\cite{arovas}, edge excitations~\cite{halperin}, and topological entanglement entropy~\cite{hamma,kitaev,levin}.

As the classic example of a gapped quantum spin liquid, we have short-ranged resonating valence bond (RVB) states originally introduced by Anderson~\cite{anderson0,anderson2,anderson,krs}, where pairs of electrons form singlet bonds and the state is a superposition of such configurations.
Rather than independently fluctuating spins, we can instead simply consider the dynamics of such valence bonds.
The low energy physics are well captured by quantum dimer models~\cite{qdmreview} (QDM) originally introduced by Rohksar and Kivelson~\cite{rk}, where the presence or absence of a dimer is indicated by an Ising degree of freedom living on the links between two sites.
The key difference between the dimer and valence bond representation being that the states corresponding to two different dimer configurations in the QDM are orthogonal by definition, but have non-zero overlap in the valence bond representation~\cite{rk}.
These models have the nice feature that at a special point, called the Rohksar-Kivelson (RK) point, the ground state can be solved for exactly and is an equal amplitude superposition of all possible dimer configurations, allowing expectation values of diagonal observables to be computed from the \emph{classical} equal probability ensemble.
The ability to describe such phases by bond variables in conjunction with a site constraint hints at a connection between such models and gauge theories. Indeed, at the microscopic level they can be formulated as hybrid lattice gauge theories with a local $U(1)$ gauge invariance~\cite{fradkin} due to the fixed number constraint at each site but with Ising valued electric fields~\cite{igt} which reflect the binary character of dimer occupations. 
The challenge in this language is to show that the gauge theory exhibits a deconfined phase which can be identified with the RVB phase.
As it turns out, the QDM on the square (or any bipartite) lattice in $d=2$ is gapless with power law decaying dimer-dimer correlations at the RK point, which sits at the boundary between a resonating plaquette~\cite{plaquettephase} and a staggered phase, and so does not host an RVB phase (upon general perturbation, one can have more complex phenomena such as Cantor deconfinement~\cite{cantordeconf}).
This lack of an RVB phase is due to the fact that the square lattice QDM maps on to a $U(1)$ gauge theory at long wavelengths~\cite{fradkin}, which is only gapless at one particular point (the RK point) in $2d$ (while there exists an extended gapless $U(1)$ RVB Coulomb phase in $3d$~\cite{3dqdm,3dqdm2,3dqdm3}).
The triangular lattice QDM, however, does exhibit exponentially decaying correlations at the RK point and hosts a fully fledged $\mathbb{Z}_2$ topologically ordered RVB liquid phase~\cite{triangular-rvb} characterized by a long wavelength $Z_2$ gauge field.  It is also useful to note that
one can also deform QDMs by loosening the fixed dimer number constraint to variable numbers. Specifically we can loosen the constraint
to allow for all odd or even numbers of dimers per site---the latter now yields a microscopic Ising gauge theory and the former its ``odd''
cousin \cite{igt}. In this limit one can find a deconfined phase on any lattice although the connection to the original RVB picture is less clear.
 [Interestingly, loosening the site constraint on the square lattice to allow one \emph{or four} dimers also allows for a deconfined $\mathbb{Z}_3$ topologically ordered phase~\cite{qdpm}].

As a natural extension of the RVB idea, the resonating singlet valence plaquette~\cite{pankov,cenke} (RSVP) generalizes from the two spin-$1/2$ $SU(2)$ singlet to $SU(n)$ singlets formed by $n$ spins in the fundamental representation of $SU(n)$ (note that the plaquette structure is not necessary, we could form $SU(n)$ singlets of $n$ spins from simplices of any form).
Following the RVB discussion, it is natural to ask whether one can find a liquid phase in these models, and if so, what is the character of this liquid?
In Ref~\onlinecite{pankov}, this idea was investigated first for $n=4$ on the simple cubic lattice, where spins formed \emph{tetramers} along the square plaquettes, with a hard-core constraint (each site was only allowed to be included in one tetramer), but was shown to exhibit a weak crystalline order (which would lead to a confining phase) at the RK point, rather than a gapped liquid~\cite{pankov,cenke}.
In fact, this current investigation was motivated by the observation that had the hard-core constraint been ``loosened'' to an even or odd constraint (that each site had to be a part of an even or odd number of tetramers), one exactly obtains the Ising ``plaquette gauge theory''~\cite{vijayhaahfu,williamson,fractonfm} in the X-Cube~\cite{vijayhaahfu} limit: a prominent example of fracton topological order~\cite{vijayhaahfu,fracton1,fracton2,fracton3,fracton4,fracton5} --- novel states of matter which exhibit quasiparticle excitations constrained to move within lower dimensional subspaces including the fracton which is a completely immobile quasiparticle.
In this context, the crystalline order at the RK point can be explained as an instability of the $U(1)$ X-cube phase to crystalline order~\cite{cenke}.
Notice how the connection between this model and the PGT parallels that of the QDM and the IGT.

This suggests that there is potentially much of interest to be found in RSVP candidates.  
In this work, we investigate another model looked at in Ref~\onlinecite{pankov}, for which Monte Carlo results show, in contrast to the cubic model,  exponentially decaying correlations at the RK point indicative of a gapped RSVP phase whose character was left undetermined.
The model is inspired from an $SU(3)$ version of the above on the face centered cubic (FCC) lattice, where three mutually nearest neighbor spins (which sit at the corners of equilateral triangular plaquettes as can be seen in Figure~\ref{fig:fccunitcell}) form an $SU(3)$ singlet.
Consequently, we may examine the quantum \emph{plaquette} model (QPM) whereby each plaquette is associated with it an Ising degree of freedom representing the presence or absence of such a singlet (a \emph{trimer}) in combination with a hard-core constraint on each site.
We describe such models in more detail in Section~\ref{sec:fccmodel}.

Given that the cubic QPM would have had Fracton order had a liquid phase existed, one might consider the possibility that the liquid phase in the FCC QPM may realize Fracton order.
Alas, this is not the case, and we show that it instead has (somewhat surprisingly) $\mathbb{Z}_3$ topological order in its liquid phase.  
This order emerges naturally from the geometry of the FCC lattice (despite the trimer degrees of freedom still being Ising), as detailed in Section~\ref{sec:hctrimer}.
Inspired by the connection between the IGT and the QDM, we examine in Section~\ref{sec:znfcc} a $\mathbb{Z}_N$ commuting-projector generalization of this model.
This model exhibits $\mathbb{Z}_3$ order when $N$ is divisible by $3$, and is trivial otherwise --- making explicitly clear the origin of the $\mathbb{Z}_3$ order in the hard-core limit.
In the Appendix, we consider similar generalizations for plaquette models on other lattices (some of which show $\mathbb{Z}_N$ fracton order).
In a sense, we make a connection between the classic ideas of RVB and RSVP and more modern concepts of topological order.  
Models with plaquette degrees of freedom have the potential to describe fracton phases (as in the simple cubic or corner-sharing octahedra lattices discussed in the appendix), or they may alternatively describe a conventional non-fracton topologically ordered phases (of which the FCC model to be discussed is an example of).

Before continuing with the discussion of the FCC QPM, we first review the key features of $\mathbb{Z}_N$ topological order in $3+1D$.~\cite{qslreview}
The theory hosts two fundamental types of excitations: a point-like quasiparticle (called the charge or ``electric'' excitation) and loop-like excitations with a finite energy per length (which we call vison~\cite{vison} loops or ``magnetic'' flux excitations).
The charge quasiparticles are self-bosons (the wavefunction does not pick up a sign upon interchanging two), but picks up a non-trivial phase when brought around a path that links with one vison loop.
More generally, bringing $n$ charge particles around a loop linked with $m$ visons result in an $e^{2\pi i n m / N}$ phase factor.
The main identifying feature of such a phase is the topological ground-state degeneracy: a system defined on a manifold with genus $g$ has an $N^g$-fold degenerate ground state that cannot be broken by local perturbations.
The different states in the ground-state manifold can be connected by the non-local action of creating a charge-anticharge pair, bringing one around the system along a non-contractible loop, and finally annihilating the pair.
We verify all these features in our model system.

\section{FCC Plaquette model}\label{sec:fccmodel}
We begin by defining a generalized plaquette model (GPM).
To clarify our nomenclature, we use ``generalized'' in the parlance of Ref~\onlinecite{rsf} to mean that we have not yet specified a site constraint.  
The quantum plaquette model (QPM) will refer specifically to the GPM with the hard-core site constraint.  
The $\mathbb{Z}_N$ generalized plaquette model examined in Section~\ref{sec:znfcc} and the Appendix will be referred to as $N$-GPM.

\begin{figure}[t]
    \centering
    \includegraphics[width=0.4\textwidth]{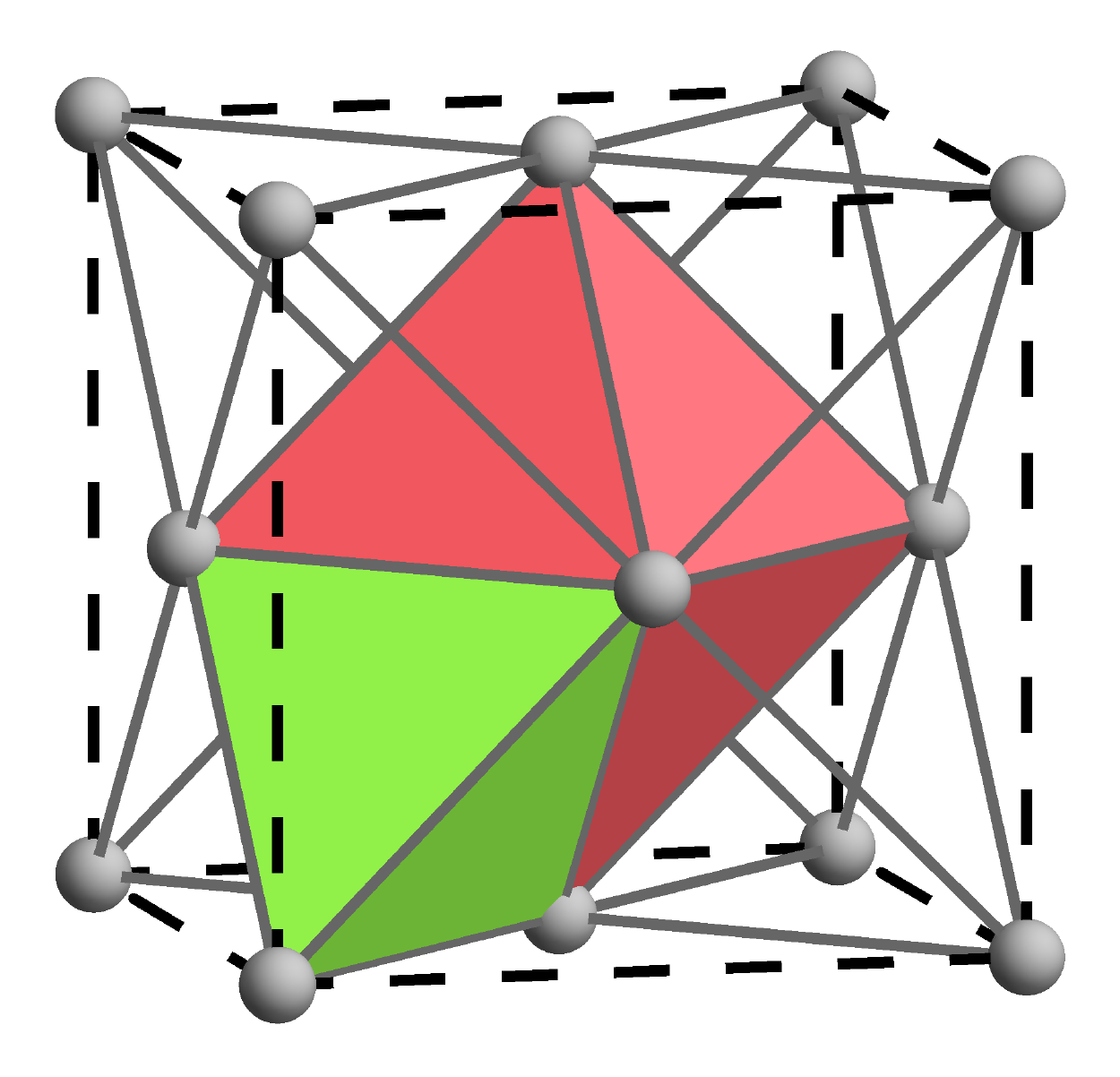}
    \caption{A unit cell of the face centered cubic lattice.  
    Nearest neighbor pairs are connected by gray lines.
    Triangles on which trimers may occupy are formed by three mutually nearest neighbor sites.
    Regular polyhedra formed by the triangular faces include octahedra (one shown in red) and tetrahedra (one shown in green).
    }
    \label{fig:fccunitcell}
\end{figure}

The model of interest is defined on the FCC lattice, a unit cell of which is shown in Figure~\ref{fig:fccunitcell}, with sites at each of the lattice points.
We will take the system defined on the 3-torus (periodic in all three directions) for simplicity.
A trimer is defined as some bound state of three mutually nearest neighbor sites, which form equilateral triangles on the FCC lattice.
We assign an Ising ($\mathbb{Z}_2$) variable $\sigma^x$ to each triangle, and define $\sigma^x=1 (-1)$ as the presence (absence) of a trimer on that triangle, and
take directly the set of all trimer configurations as an orthonormal basis for our Hilbert space.

We may now begin to discuss Hamiltonians on this Hilbert space.
These will consist generically of three parts: a site constraint, a kinetic term, and a potential term.
The site constraint is a local constraint diagonal in the trimer basis, which is defined for each site and must be satisfied, thus permitting only a subset of the Hilbert space.  
This constraint may be enforced externally, or energetically on the ground state by attaching a large energy penalty to violating states.
For example, the QPM will be obtained by enforcing that each site is only allowed to be a part of exactly one trimer, but one can also write down a theory where each site is only allowed to be a part of an odd (even) number of trimers (thus producing a kind of Ising ``plaquette gauge theory'').
The kinetic term is a sum of purely off-diagonal local terms that transition between trimer configurations respecting by the site constraint.
Finally, the potential term is a sum of diagonal local terms, which may be used to tune the Hamiltonian to the RK point --- where the ground state can be solved for exactly!

Before jumping straight to the hard-core QPM, one might expect that there may be something to learn first from the GPM with the odd/even constraint.
This expectation turns out to be wrong: the exactly solvable even/odd models are actually non-topologically ordered liquids.  
First, note that the even and odd models are unitarily related, thus it is only necessary to examine the even case.
Let us write this down explicitly for the even model.
The Hamiltonian is given by
\footnote{Note that this even model is also what one would have found starting from an FCC Ising model with triangular plaquette interactions and proceeded with the generalized gauging procedure, or FS-duality, of Ref~\onlinecite{vijayhaahfu}.}
\begin{equation}
    \mathcal{H}_{\text{even}} = -\sum_{\mathcal{C}_e}\prod_{t\in \mathcal{C}_{e}} \sigma^z_t - \sum_{s}\prod_{t\in s}\sigma^x_t
    \label{eq:Heven}
\end{equation}
where $t$ refers to triangles, and $\sigma^{z,x}_t$ are Pauli matrices acting on the trimer degree of freedom on each triangle.  The second sum is over sites $s$, and $t\in s$ corresponds to the triangles containing the site $s$ (of which there are 24 of).  
The set $\mathcal{C}_{e}$ refers to a set of triangles for which each site on the lattice is shared by an even number of triangles in $\mathcal{C}_{e}$ (thus guaranteeing the term commutes with the site constraint), and $\mathcal{C}_{e}$ does not consist of multiple disjoint sets of triangles (the subscript $e$ stands for \emph{even}).  
The first sum is over all such sets $\mathcal{C}_{e}$ up to a certain size ${|\mathcal{C}_{e}|}_\text{max}$, which we will assume is large enough for ergodicity (within a topological sector, should they exist).
We will return to the discussion of what these terms look like in more detail in the context of the (hard-core) QPM in Sec~\ref{sec:hctrimer}.
The first term is the kinetic term, and the second term enforces the constraint that every site must have an even number of trimers connected to it (there is no potential term needed here).
By construction, this Hamiltonian consists of mutually commuting terms and one can deduce that an equal amplitude superposition of all constraint-satisfying configurations within a topological sector (should they exist) is the exact ground state.

In fact, \emph{no such topological sector exists}.  An easy way to see this is by examining the excitation structure.
In the gauge theory language, consider creating a ``charge'' excitation: an excitation of the second term in the Hamiltonian, where a site participates in an odd number of trimers.
It is in fact possible to create a single such an excitation locally at site $s$ by applying an operator $\sigma^z_{t_1}\sigma^z_{t_2}\sigma^z_{t_3}$ on the ground state, where $t_1,t_2,t_3$ are the three triangles around a tetrahedron that contain the site $s$.
These overlap the site $s$ three times, and the three other sites in the tetrahedron twice, thus it anticommutes with the site term only on site $s$.
We have therefore created a single charge excitation using only local operators acting on the ground state, thus implying that a single charge excitation \emph{does not carry any topological charge}.
By topological charge, we refer to charge that can be measured by a membrane-like operator akin to Gauss' law in standard $U(1)$ electromagnetism.
We are therefore forced to conclude that this Hamiltonian does not possess the features of topological order such as topological degeneracy and quasiparticle/loop excitations with non-trivial statistics.
Nevertheless, as we will show in the next section, the QPM (specified by a \emph{number} site constraint) at the RK point \emph{does} exhibit the signs of topological order, more specifically, $\mathbb{Z}_3$ topological order.
The reason the above construction fails is that we have implicitly tried to force a $\mathbb{Z}_2$ order by using an even constraint, while the \emph{geometry} of the model favors a $\mathbb{Z}_3$ order.

\section{The Hard-Core constraint}\label{sec:hctrimer}
We now examine the FCC QPM: the model of trimers with the hard-core constraint that each site must participate in only one trimer.
The allowed Hilbert space now consists of the set of hard-core trimer coverings of the FCC lattice.
The set of local trimer moves are now more restricted than in the even theory.  
Any local trimer move can be represented by a non-disjoint bipartite set of triangles $\mathcal{C}=\mathcal{C}_A\cup\mathcal{C}_B$, with the constraint that every site in the lattice must be included in exactly one triangle from $\mathcal{C}_A$ and one from $\mathcal{C}_B$, or none at all.  
By non-disjoint, we mean that one cannot express $\mathcal{C}$ as $\mathcal{C}=\mathcal{C}_1\cup\mathcal{C}_2$ for $\mathcal{C}_{1,2}$ both being valid bipartite sets as previously defined.
The trimer move then consists of taking all trimers that were originally on all the triangles in $\mathcal{C}_A$ and moving them to $\mathcal{C}_B$, or vice versa.
Let us represent the local state in which all triangles in $\mathcal{C}_A$ are occupied with trimers as $|\mathcal{C}_A\rangle$, and similarly $|\mathcal{C}_B\rangle$.
We can then define a RK type model as
\begin{eqnarray}
    \mathcal{H}_{RK} = -t\sum_{\mathcal{C}} \left(|\mathcal{C}_A\rangle\langle\mathcal{C}_B|+|\mathcal{C}_B\rangle\langle\mathcal{C}_A|\right)\\
    +V\sum_{\mathcal{C}}\left(|\mathcal{C}_A\rangle\langle\mathcal{C}_A| + |\mathcal{C}_B\rangle\langle\mathcal{C}_B|\right)\nonumber
\end{eqnarray}
where the sum is over all $\mathcal{C}$ as previously described up to some $|\mathcal{C}|_\text{max}$.
We further have the site constraint of one trimer per site: $\sum_{t\in s}(\sigma_t^x+1)/2 = 1$ for every site $s$.
This can be expressed as enforcing the constraint $G_s|\psi\rangle = |\psi\rangle$ for all $s$ with
\begin{equation}
    G_s = e^{-i\alpha\left[1-\sum_{t\in s}(\sigma_t^x+1)/2\right]}\label{eq:1constraint}
\end{equation}
for \emph{any} $\alpha$.
Note that this Hamiltonian, written in terms of Pauli matrices, has a $U(1)$ symmetry $\sigma^{\pm}_t \rightarrow e^{\pm i \alpha}\sigma^{\pm}_t$, where $\sigma^{\pm}=\sigma^y\pm\sigma^z$ are $\sigma^x$ raising/lowering operators.
This $U(1)$ symmetry corresponds to the conservation of total trimer number, as every such bipartite path satisfies $|\mathcal{C}_A|=|\mathcal{C}_B|$.

Exactly at $t=V$, the RK point, the Hamiltonian is a sum of projectors,
\begin{eqnarray}
    \mathcal{H}_{RK}^{t=V=1} = 2\sum_\mathcal{C} (|\mathcal{C}_A\rangle- |\mathcal{C}_B\rangle)
    (\langle\mathcal{C}_A|- \langle\mathcal{C}_B|)
    \label{eq:fccrk}
\end{eqnarray}
whose exact ground state is an equal amplitude sum of all constraint-obeying trimer configurations that can be reached by the local flips $\mathcal{C}$.

At the RK point, which will be the focus of our discussion, expectation values of diagonal operators are exactly that of the equal probability classical ensemble.
The trimer-trimer correlation function at the RK point was calculated via Monte Carlo simulation in Ref~\onlinecite{pankov}, and was found to decay exponentially with a small correlation length.
This indicates that should a suitable RK type Hamiltonian be defined, the RK point sits within a gapped RSVP phase 
--- if the RK point were a critical point between two phases or part of a gapless phase, one would expect power law decaying correlations (another unlikely scenario is the existence of two first-order transitions directly on either side of the RK point, which we do not consider).

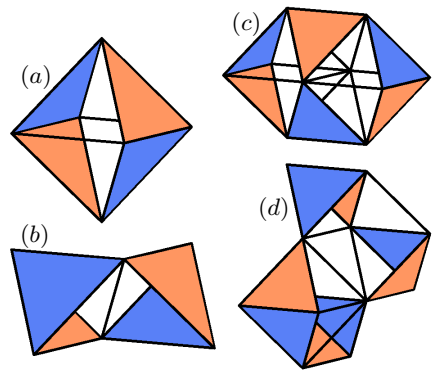
\begin{figure}[t]
    \centering
    \begin{picture}(150,150)
    \put(-20,0){\includegraphics[width=0.4\textwidth]{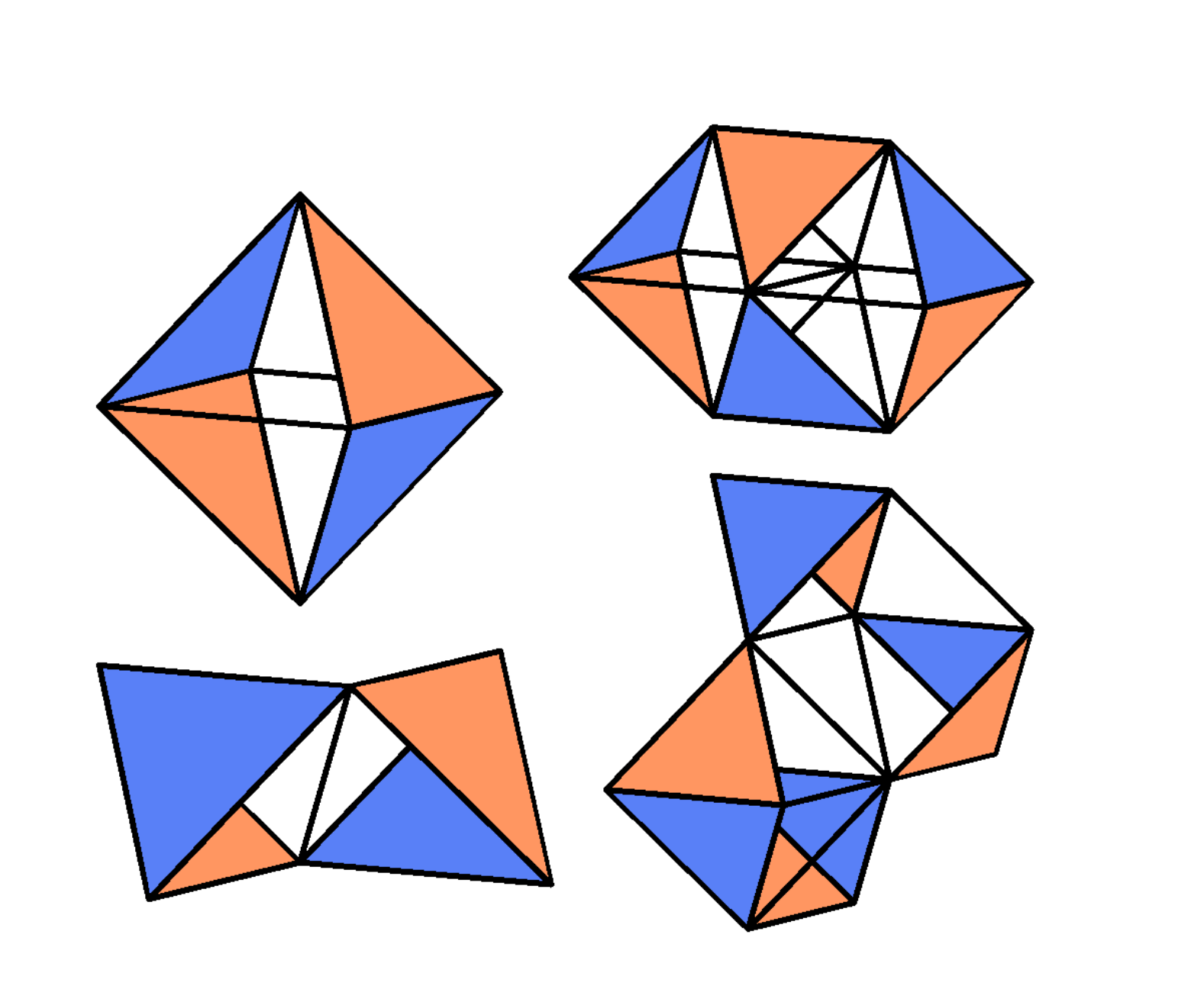}}
    \put(-00,120){$(a)$}
    \put(-00,60){$(b)$}
    \put(80,140){$(c)$}
    \put(90,70){$(d)$}
    \end{picture}
    \caption{Illustration of a few terms in the Hamiltonian, which we describe by sets of triangles $\mathcal{C}=\mathcal{C}_A\cup\mathcal{C}_B$, where the orange and blue triangles indicate $\mathcal{C}_A$ and $\mathcal{C}_B$.
    All $|\mathcal{C}|=4$ terms are loop terms of the form $(a)$ or $(b)$.
    $(c)$ and $(d)$ shows terms involving a larger number of triangles.
    The term $(c)$ involves flipping between configurations with local ``divergence'' $\pm 3$ (as described in the text),
    and $(d)$ is an example of a $|\mathcal{C}|=8$ length loop term.
    }\label{fig:terms}
\end{figure}

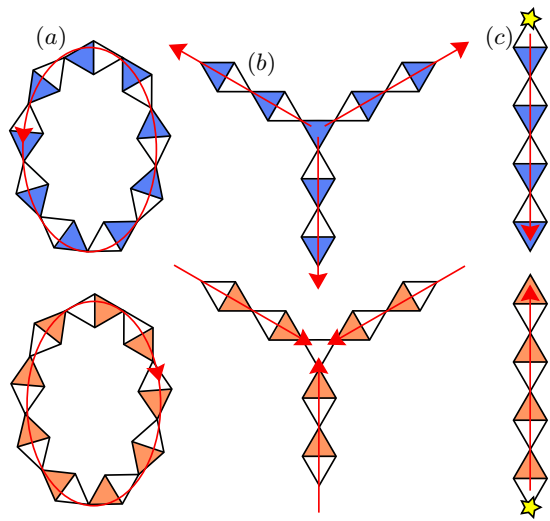
\begin{figure}[t]
    \centering
    \begin{picture}(150,200)
    \put(-30,0){\includegraphics[width=0.4\textwidth]{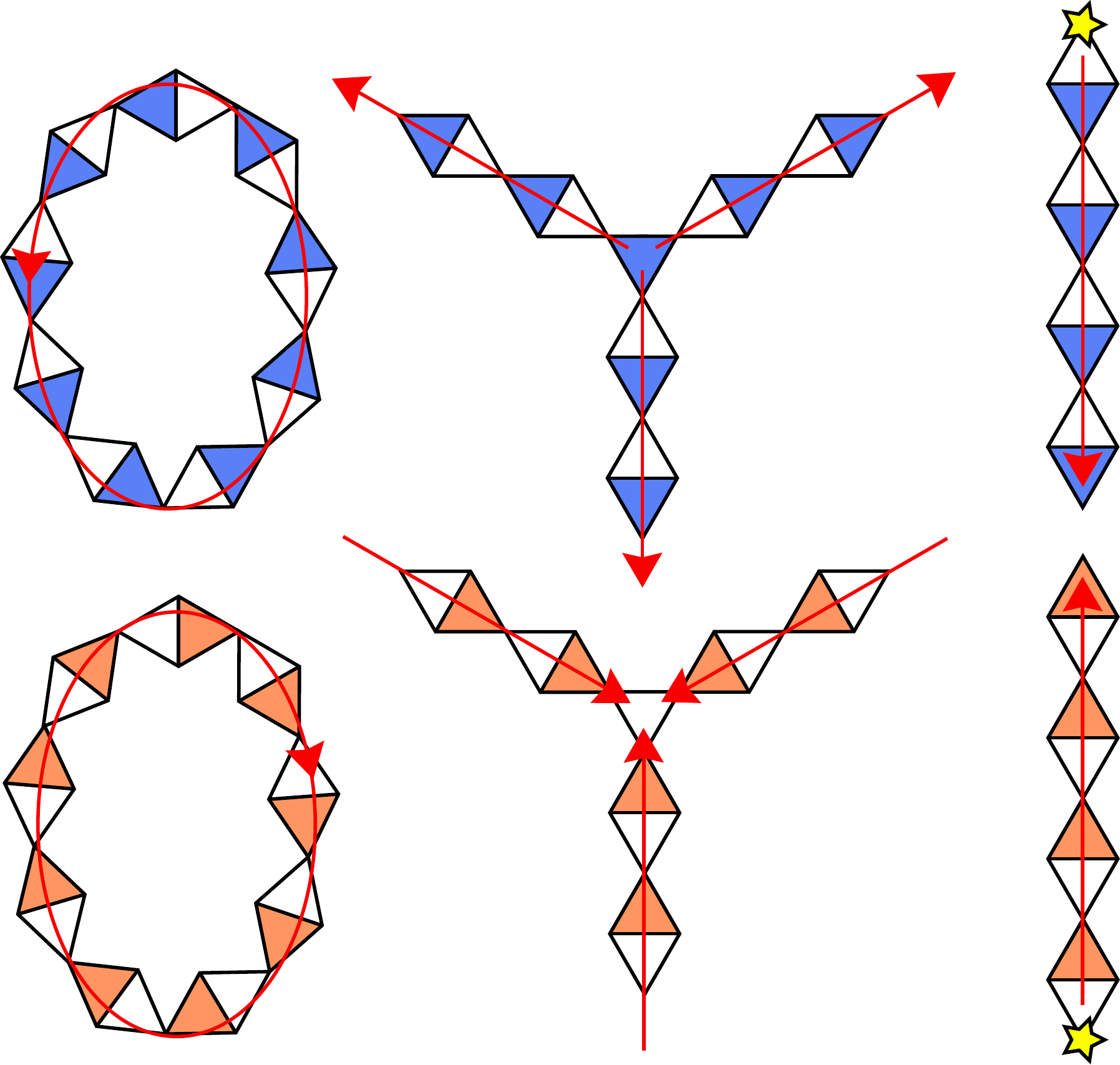}}
    \put(-20,180){$(a)$}
    \put(60,170){$(b)$}
    \put(150,180){$(c)$}
    \end{picture}
    \caption{The convention for assigning directions to trimer configurations.
    The top row shows the configuration (for example) in state $|\mathcal{C}_A\rangle$, and the bottom shows the flipped state $|\mathcal{C}_B\rangle$; the red arrows indicate the direction assignment.
    Configurations along loop-like paths are assigned a direction as shown in $(a)$.
    Terms which involve flips along non-loop paths include triangles with local ``divergence'' $\pm3$, as shown in $(b)$.
    Finally, $(c)$ shows how a monomer (an untrimerized site) may be moved along a path via trimer flips.
    }\label{fig:direction}
\end{figure}

Let us now discuss what possible terms, denoted by the set of flipped triangles $\mathcal{C}$, arise in our model and how large clusters $|\mathcal{C}|_\text{max}$ one should include for ergodicity.
The simplest types of moves are loop moves, where $\mathcal{C}$ consists of a loop of an even number of triangles joined in alternating orientation (each triangle shares sites with only two other triangles, as shown in Figure~\ref{fig:direction}a). 
The smallest moves are $|\mathcal{C}|=4$ terms of this type, which come in two flavors: a loop of four triangles around an octahedron, and a loop of four triangles around two edge-sharing tetrahedra, shown in Figure~\ref{fig:terms}a and \ref{fig:terms}b.

To more effectively visualize the action of these loop terms, we can unambiguously assign a \emph{directionality} to the loop configurations $|\mathcal{C}_A\rangle$ and $|\mathcal{C}_B\rangle$.  
To set a convention, imagine the triangles in $\mathcal{C}_A$ as arrowheads which all point in one direction around the loop, which we define to be the direction of the configuration $|\mathcal{C}_A\rangle$, as shown in Figure~\ref{fig:direction}a.  
Similarly, we may look at the configuration $|\mathcal{C}_B\rangle$, which always points in the opposite direction.
Pictorially, the kinetic term then looks like $-t(|\hspace{-1.0mm}\circlearrowleft\rangle\langle \circlearrowright\hspace{-1.0mm}| + |\hspace{-1.0mm}\circlearrowright\rangle \langle \circlearrowleft\hspace{-1.0mm}|)$ in this language.
In this description, the loop terms are always flipping between ``divergenceless'' configurations.
A flip is characterized as a loop if every triangle is only in contact with two other triangles.  
However, a triangle may also be in contact with \emph{three} other triangles.
In our picture, such triangles have a ``divergence'' of $\pm3$, as shown in Figure~\ref{fig:direction}b.
Terms involving such triangles first appear in the Hamiltonian at $|\mathcal{C}|=6$, one such example is shown in Figure~\ref{fig:terms}c.

As we will show, there exists a conserved number that is left invariant under local trimer manipulations, modulo 3.
However, the loop terms with $|\mathcal{C}|=4$ leave this number unchanged \emph{not} modulo 3 and we have an extra unwanted conservation law that we can get rid of by including larger terms.
At $|\mathcal{C}|=6$, the term in Figure~\ref{fig:terms}c is sufficient to accomplish this, and at $|\mathcal{C}|=8$, there are larger loop terms such as the one shown in Figure~\ref{fig:terms}d that also accomplish this.
Thus, we need \emph{at least} $|\mathcal{C}|_\text{max}=6$ to achieve ergodicity.  
We do not investigate this question of ergodicity further here, and assume that there is a small finite value of $|\mathcal{C}|_\text{max}$ (which may just be 6) for which the Hamiltonian is ergodic \emph{enough} within each topological sector.

\begin{figure}[t]
    \centering
    \includegraphics[width=0.4\textwidth]{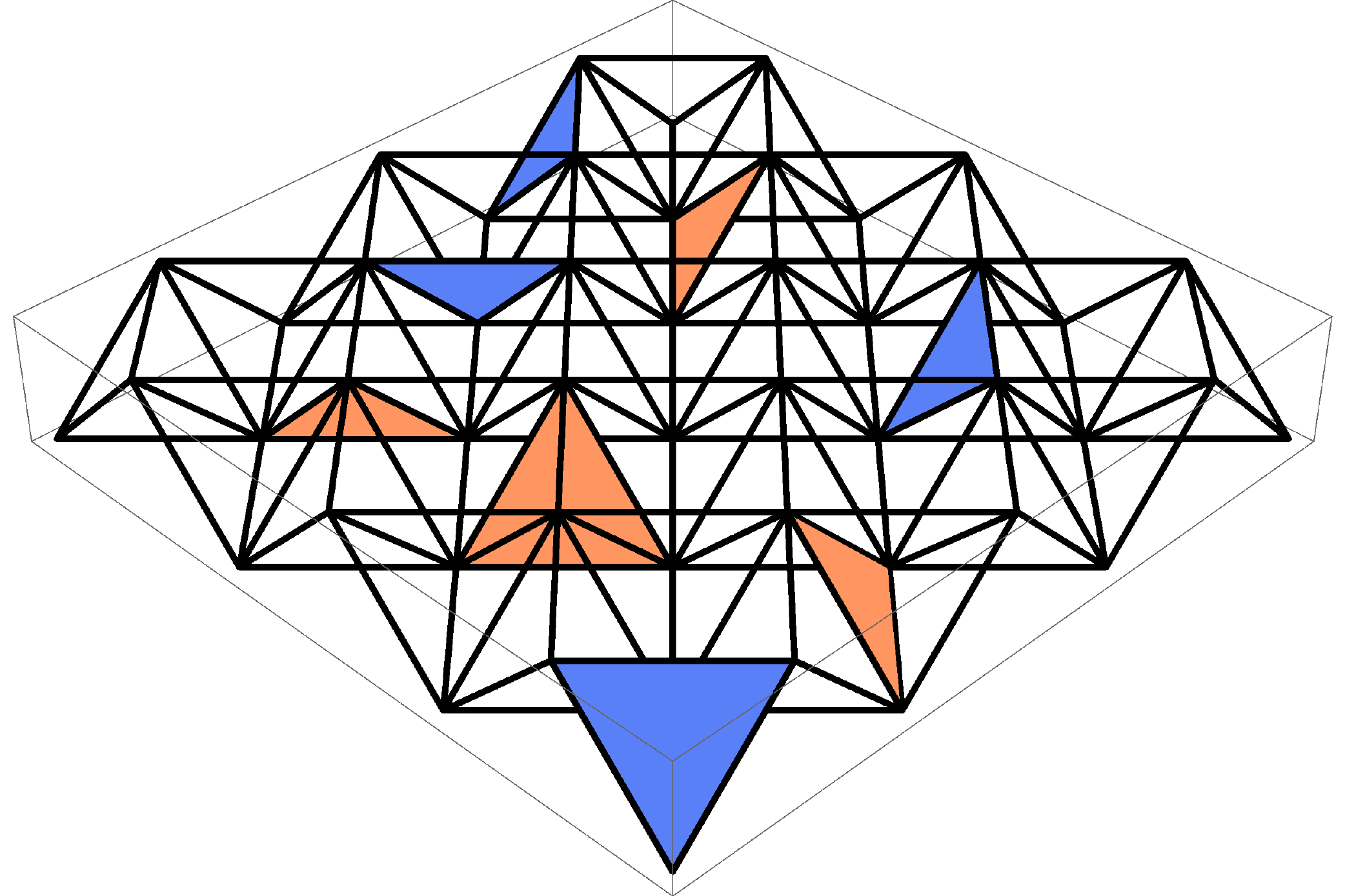}
    \caption{A sample trimer configuration in an $xy$ plane specified by $z$-coordinate $z_0$, which includes triangles spanning the site-layers $z_0$ and $z_0+1/2$.
    Upwards facing trimers are shown in orange, while downwards facing trimers are shown in blue.
    The topologically conserved ``winding number'' is the difference between the number of upwards facing trimers ($N_\triangle$) and downwards facing trimers ($N_\triangledown$) modulo 3 (Eq~\ref{eq:winding}).
    }\label{fig:conserved}
\end{figure}

We can now proceed to discuss conserved quantities that remain invariant under such local flips.
Consider two adjacent $xy$-plane of sites defined by the $z$-coordinate $z_0$ and $z_0+1/2$ of the FCC lattice, as shown in Figure~\ref{fig:conserved} (where the linear dimension of the cubic unit cell is taken to be $1$).
All the triangles with all three sites within these two planes are oriented with either: two sites on the lower and one on the upper, which we call ``upwards pointing'' ($\triangle$), or the opposite, which we call ``downwards pointing'' triangles ($\triangledown$).
We claim that the ``winding number'' for this $xy$ plane,
\begin{equation}
 W_{xy}^{(z_0)}=N_{\triangle}^{(z_0)}-N_{\triangledown}^{(z_0)}\mod 3   
 \label{eq:winding}
\end{equation}
 is conserved by arbitrary local trimer moves, where $N_{\triangle}^{(z_0)} (N_{\triangledown}^{(z_0)})$ is the number of upwards (downwards) pointing trimers between layers $z_0$ and $z_0+1/2$.

Furthermore, knowing $W_{xy}^{(z_0)}$ for one $z_0$ determines the value for all other $xy$ planes.
We can show this using a simple counting argument.
The number of sites on layer $z_0+1/2$ that are included in the trimers spanning $z_0, z_0+1/2$, is $N_{\triangle}^{(z_0)}+2N_{\triangledown}^{(z_0)}$.  
Let $N_{xy}$ be the total number of sites an $xy$ layer.
This leaves $N_{xy}-(N_{\triangle}^{(z_0)}+2N_{\triangledown}^{(z_0)})$ free sites in layer $z_0+1/2$ that must be used in the trimers spanning $z_0+1/2,z_0+1$, as there are no untrimerized (monomer) sites.
Therefore, we must have
\begin{equation}
 2N_{\triangle}^{(z_0+1/2)}+N_{\triangledown}^{(z_0+1/2)} =    N_{xy}-(N_{\triangle}^{(z_0)}+2N_{\triangledown}^{(z_0)}) 
    ,
\end{equation}
and taking both sides modulo 3, we find
\begin{equation}
     W_{xy}^{(z_0+1/2)} = W_{xy}^{(z_0)} - N_{xy}  \mod 3.
\end{equation}
Therefore, knowing $W_{xy}^{(z_0)}$ for $z_0$ fixes its value for every $z$.
This alone is proof that $W_{xy}^{(z_0)}$ cannot be modified by any local trimer move: to modify one we must simultaneously change this value for \emph{every} value of $z$, which requires a non-local trimer move.
The same argument holds for the $yz$ and $zx$ planes, which therefore give us access to three independent conserved winding numbers.
Measuring these winding numbers requires counting the number of triangles within an entire plane: a non-local measurement. 
At the RK point (and the RSVP liquid phase), this leads to a locally indistinguishable $3^3$-fold degenerate ground state manifold on a 3-torus.
Thus, we have already uncovered the topological ground state degeneracy --- a key features of a $\mathbb{Z}_3$ topologically ordered phase.

Next, we observe that the non-local trimer shift needed to change these winding numbers correspond to flips on paths $\mathcal{C}$ that are equivalent to non-contractible loops.
Consider the non-local trimer loop move $\mathcal{C}$ which runs along a non-contractible loop wrapping once around the $z$ direction.
Let $|\mathcal{C}_A\rangle$ be the configuration where the ``direction'' of the loop as previously discussed points along the positive $z$ direction, and $|\mathcal{C}_B\rangle$ along the negative direction.
Then, flipping $|\mathcal{C}_B\rangle\rightarrow|\mathcal{C}_A\rangle$ will increment $W_{xy}^{(z_0)}$ by 1.
Since $W_{xy}^{(z_0)}$ for every slice must be changed identically, we further see that any further local manipulations one makes to the details of  $\mathcal{C}$ will not change its effect on $W_{xy}^{(z_0)}$.

To complete the picture of the $\mathbb{Z}_3$ topological order, we next consider the form of the excitations. 
At the RK point, we only have the ground state that can be solved for exactly --- and while we can write down variation states with localized excitations, these will not be exact (they must be locally ``dressed'' and the true eigenstates will be a definite momentum superposition)~\cite{rk}.
We examine two types of excitations in this model:
point-like monomer (``charge'') excitations and loop-like vison (``magnetic'') excitations.

Monomer excitations are sites which do not participate in any trimer.  To include these, we must relax our constraint in Eq.~\ref{eq:1constraint} to allow states with $G_s|\psi\rangle=e^{i\alpha}|\psi\rangle$ at some energy cost.
A single monomer can be moved from site $s$ to $s^\prime$ by a trimer flip along a path, as shown in Figure~\ref{fig:direction}c.
Adding a two-triangle hopping term gives monomer excitations a finite mass and dispersion.
We can now identify the non-local flip that increments the winding number by one as corresponding to the action of bringing a monomer excitation around along a non-contractible loop in the negative $z$ direction once.

To create vison excitations, consider a loop $\mathcal{L}$, and let $W_{\mathcal{L}}$ count the winding number as previously defined in Eq~\ref{eq:winding} but for an open surface with boundary at $\mathcal{L}$.
We then define the ``vison  operator'' as 
    $v_{\mathcal{L}} = e^{2\pi i W_{\mathcal{L}}/3}$.
Our cartoon state containing a vison loop along $\mathcal{L}$ will then look like 
\begin{equation}
    |v_\mathcal{L}\rangle \approx |W_\mathcal{L}=0\rangle+e^{2\pi i/3}|W_\mathcal{L}=1\rangle + e^{-2\pi i/3}|W_\mathcal{L}=2\rangle
\end{equation}
where $|W_\mathcal{L}=k\rangle$ is the component of the ground state wavefunction with $W_\mathcal{L}=k$.
Any term in the Hamiltonian far away from the loop $\mathcal{L}$ does not change the value of $W_\mathcal{L}$, and so this state remains a local eigenstate of those terms.  
This is not true for terms near the loop which do change the value of $W_\mathcal{L}$, and so this state will have a finite energy density along $\mathcal{L}$ (but will not be an eigenstate of those terms).
In this cartoon picture, one can imagine threading $n$ monomer excitations through $m$ vison loops before returning to its original position, resulting in an overall phase $e^{2\pi i n m/3}$ (of course, actually rigorously defining such a process requires more care).

Thus, we have shown that the QPM in its RSVP phase does indeed possess $\mathbb{Z}_3$ topological order, with all of its important features.
In the next section, we will examine a $\mathbb{Z}_N$ generalization of the FCC QPM in an exactly solvable limit, which shares much of the properties of the hard-core model just discussed, including a $\mathbb{Z}_3$ order \emph{for all} $N$ divisible by 3.
The properties of these models generically depend strongly on $N$ and the details of the lattice, and for the interested reader we cover a few more characteristic examples in the Appendix.

\section{$\mathbb{Z}_N$ Generalization}\label{sec:znfcc}
To motivate the study of the $\mathbb{Z}_N$ generalization, we observe that by doing a simple operator substitution on the hard-core Hamiltonian, one can get a Hamiltonian of mutually commuting projectors which can be solved exactly.

The first step is to enlarge the $\mathbb{Z}_2$ degree of freedom on each plaquette to a $\mathbb{Z}_N$ degree of freedom.
Acting on each of these degrees of freedom, we have the operators $X$,$Z$, for each bond obeying algebra
\begin{eqnarray}
    Z^N = X^N = 1\nonumber\\
    XZ = \omega ZX
    \label{eq:znalg}
\end{eqnarray}
where $\omega=e^{2\pi i/N}$.
Thus, the eigenvalues of $X$ are $\omega^n$ for $n=1\dots N$, and $Z$ acts as a raising operator in the $X$ eigenbasis.
Interpreting the $X$ eigenvalue $\omega^n$ as the presence of $n$ trimers on a bond, we can then enforce a site constraint that the sum of trimers connected to a site always be zero mod $N$.  
For large $N$, these can be interpreted as bosonic or quantum rotor degrees of freedom, as in Ref~\onlinecite{motrunich,motrunich2}.
Note that we could have equally chosen the site constraint to be any number without changing the physics, as the resulting Hamiltonians can be shown to be unitarily related to each other.
Quantum dynamics that respect this constraint can then be represented by substituting $\sigma^+\rightarrow Z, \sigma^-\rightarrow Z^\dagger$ in the kinetic term of the RK Hamiltonian Eq~\ref{eq:fccrk} when expressed in terms of raising/lowering operators.
Since the kinetic term does not annihilate any state, the potential term is not needed.

Thus, we have 
\begin{eqnarray}
    \mathcal{H}_{N} &=& -\sum_{\mathcal{C}}\left(\prod_{t\in\mathcal{C}_A}Z_t\prod_{t\in\mathcal{C}_B}Z_t^\dagger + h.c.\right)\\
    &&- \sum_{s} \left(\prod_{t\in s}X_t + h.c.\right)\nonumber
\end{eqnarray}
where the first sum is over all bipartite connected sets of triangles $\mathcal{C}=\mathcal{C}_A\cup\mathcal{C}_B$ such that every site contains an equal number of triangles from $\mathcal{C}_A$ and $\mathcal{C}_B$. 
Note that this is a looser constraint than in the hard-core case (where each site had to have \emph{one} from each, or none).

We can motivate that this model will have $\mathbb{Z}_3$ order only if $N$ is a multiple of 3, and trivial otherwise, by just looking at the quasiparticle structure.  
We may define the charge as $Q_s = \prod_{t\in s} X_t$, where the product is over the 24 triangles touching a site.
However, acting with $Z_t$ creates a set of three charges $\omega$ each, and so we are therefore forced to conclude that three charges combined carries no topological charge
(note that if the lattice were tripartite, then a different charge definition could be used on each sublattice and this conclusion would not hold --- some examples of this happening are discussed in the Appendix).
If $N$ is not a multiple of three, then one can create a single $\omega$ charge via local operations, and we are left with a trivial liquid.
On the other hand, if $N$ is a multiple of three, there is the possibility for a $\mathbb{Z}_3$ topological order.
In this situation, the correct definition of the topological charge operator should be 
\begin{equation}
    Q^\text{top}_s = Q_s^{N/3}.
\end{equation}
We assume that $N$ is a multiple of three moving forwards.

First, note that there may be non-topological degeneracies that exist due to commuting terms which are not included in the Hamiltonian because they cannot be expressed as products of terms on bipartite $\mathcal{C}$.  
The product ${(Z_{t_1}Z_{t_2}Z_{t_3}Z_{t_4})}^{N/3}$ around the four faces of a tetrahedron is such an example, which leads to an extra 3-fold non-topological degeneracy.
We will ignore non-topological degeneracies as they can be broken by local perturbations.

To count the topological degeneracy, 
consider the operator that counts $N_{\triangle}^{(z_0)}-N_{\triangledown}^{(z_0)}$ for an $xy$-plane of triangles (as considered earlier for QPM),
\begin{equation}
    e^{2\pi i (N_\triangle^{(z_0)} - N_\triangledown^{(z_0)})/N} = \prod_{t\in\triangle}X_t \prod_{t\in\triangledown}X_t^\dagger
\end{equation}
where the product $t\in\triangle$ ($t\in\triangledown$) is over all upwards (downwards) pointing triangles in the $xy$-plane spanning $z_0$,$z_0+1/2$.
While this commutes with all $|\mathcal{C}|=4$ terms in the Hamiltonian, it fails to do so with some $|\mathcal{C}|=6$ terms, (such as the one shown in Figure~\ref{fig:terms} for the QPM), and general local perturbations.
Instead, like in the QPM, this number is only conserved mod $3$ under local operations, and so the correct operator is 
\begin{equation}
    W_{xy} = {\left(\prod_{t\in\triangle} X_t \prod_{t\in\triangledown}X_t^{\dagger}\right)}^{N/3}
\end{equation}
which \emph{does} commute with every term in the Hamiltonian.
We have suppressed the $z_0$ label, as it is possible to relate $W_{xy}^{(z_0)}$ for different $z$ by terms present in the Hamiltonian.
To see this, observe that multiplying $W_{xy}^{(z_0)}$ by ${(Q^{\text{top}}_s)}^\dagger$ on every site $s$ in the $z_0+1/2$ layer results in $W_{xy}^{(z_0+1/2)}$, and so therefore $W_{xy}^{(z_0)} = W_{xy}^{(z_0+1/2)}$ in the ground state where $Q_s^\text{top} = 1$. 
We have $W_{xy}^3=1$ and so $W_{xy}$ can take on one of three values, and since there are three independent planes one could have defined this for, this leads to a $3^{3}$ topological degeneracy.
Notice the remarkable similarity to the QPM discussion in Section~\ref{sec:hctrimer}.

The advantage of this model over the QPM at the RK point is that the excitations are static can be solved exactly.
A monomer excitation from the QPM correspond to a $Q_s=\omega$ charge sitting on a site $s$, which carries topological charge $Q_s^\text{top} = e^{2\pi i/3}$.
By application of a chain operator $Z_{t_1}^\dagger Z_{t_2} \dots Z_{t_{L-1}}^\dagger Z_{t_L}$, a monomer can be moved from one site to another, and moving one monomer around a non-contractible loop in the $z$ direction will modify the value of the conserved winding number $W_{xy}$ by $e^{\pm 2\pi i /3}$ depending on which direction the monomer goes around the loop.

The vison (magnetic) excitations of this model are loop-like, and are created at the boundary of a membrane operator,
\begin{equation}
    W_{\mathcal{L}} = {\left(\prod_{t\in\triangle^{\mathcal{L}}} X_t\prod_{t\in\triangledown^{\mathcal{L}}} X_t^\dagger\right)}^{N/3}
\end{equation}
where $\triangle^\mathcal{L}$ ($\triangledown^\mathcal{L}$) are all the upwards (downwards) oriented triangles along an open surface with boundary along the loop $\mathcal{L}$ (which we may take to be a flat loop in an $xy$ plane, where this operator can be thought of as a truncated version of the $W_{xy}^{(z_0)}$ operator).
Acting with this operator on the ground state creates an excited eigenstate of the Hamiltonian, which is locally the ground state away from $\mathcal{L}$, but an excited eigenstate with gap $2(1-\cos{2\pi /3})$ for each term near the loop $\mathcal{L}$ that doesn't commute with $W_\mathcal{L}$.

We can now also explicitly verify the statistical phase obtained by bringing charge excitations through vison loops.
Consider the action of bringing $n$ charge excitations around in a circle linking with $m$ vison loops, bringing us back to the same state but with a overall phase.
In the simplest case, computing this phase involves commuting a $Z^n$ with $(X^{\dagger})^{N m/3}$, which results in a $\omega^{N n m / 3} = e^{2\pi i n m / 3}$ phase factor overall, in agreement with what one expects from a $\mathbb{Z}_3$ phase.

Finally, we note that such a $\mathbb{Z}_N$ model can in principle be defined on any lattice, and produces a wide variety of interesting topological phases.
We have examined a few characteristic cases in the Appendix. 

\section{Conclusion}

To conclude, we have investigated in detail the topological properties of the FCC QPM, a prime candidate for an RSVP phase.
In doing so, we discovered the presence of a $\mathbb{Z}_3$ topological conserved quantity that leads to a $3^3$-fold topological ground state degeneracy at the RK point on a $3$-torus, where
 this model was shown to have exponentially decaying trimer-trimer correlations~\cite{pankov} indicating the presence of a gapped liquid RSVP phase.
Our result would then imply that this topological degeneracy is a feature of the whole phase, and we show that it also shares the features one expects of a phase can be described by a $\mathbb{Z}_3$ gauge theory, such as $\mathbb{Z}_3$ quasiparticle excitations and loop-like vison excitations.
This $\mathbb{Z}_3$ emerges naturally from the \emph{geometry} of the FCC lattice, in the same way that a $\mathbb{Z}_2$ order emerges in the triangular lattice QDM.

\acknowledgements
TD thanks Shivaji Sondhi for suggesting this model be studied in the first place and for continued guidance and support on this project.
TD also thanks Roderich Moessner, Siddharth Parameswaran, Daniel Arovas, Barry Bradlyn, Sanjay Moudgalya, and Christian Jepsen for many helpful discussions.

\appendix

\section{$\mathbb{Z}_N$ Generalized Models on other lattices}\label{sec:zn}

We motivate the study of these $\mathbb{Z}_N$ generalized models from an observation that by doing a simple operator substitution on the hard-core Hamiltonian for QDMs or QPMs, one gets a Hamiltonian of mutually commuting projectors which can be solved exactly.
Some possible phases found in these exactly solvable models are summarized in Table~\ref{tab:znmodels}.
We will refer to such models as $N$-GDM (specifically for dimer models), and $N$-GPM for the plaquette models (which include trimer models and a tetramer model which we also discuss).

\begin{table}[t]
    \centering
    \begin{tabular}{r|l |l|}
    $\mathbb{Z}_N$ model & Lattice & Phase\\
    \hline
    \multirow{2}{*}{Dimer} & Square & $\mathbb{Z}_N$ \\
    \cline{2-3}
    & Triangular & $\mathbb{Z}_{\gcd(2,N)}$\\
    \hline
    \multirow{4}{*}{Trimer} & Triangular& $\mathbb{Z}_N\times \mathbb{Z}_N$\\
    \cline{2-3}
    & Corner-sharing  &   $\mathbb{Z}_N$ fracton \\
    &   octahedra     &    (X-cube phenomenology)          \\
    \cline{2-3}
    & Face centered cubic & $\mathbb{Z}_{\gcd(3,N)}$\\
    \hline
    \multirow{1}{*}{Tetramer} & Simple Cubic& $\mathbb{Z}_N$ fracton (X-cube)\\
    \hline
    \end{tabular}
    \caption{Table summarizing the topological phases found for the $\mathbb{Z}_N$ generalized dimer models (first two rows) and $\mathbb{Z}_N$ generalized plaquette models (remaining rows).
    $\mathbb{Z}_{\gcd(p,N)}$ for $p=2,3$ simply means $\mathbb{Z}_p$ order if $N$ is a multiple of $p$, and trivial otherwise.
    The FCC QPM is discussed in Section~\ref{sec:hctrimer} of the main text.
    }
    \label{tab:znmodels}
\end{table}

To illustrate the construction for a general lattice model, we first consider the Rohksar-Kivelson QDM on the square lattice. 
Letting $\sigma^x=1 (-1)$ on a bond signify the presence (absence) of a dimer, we can write the Hamiltonian as
\begin{eqnarray}
    \mathcal{H}_\text{RK} &=& -t \sum_{\square}\sigma^+_{l_1}\sigma^-_{l_2}\sigma^+_{l_3}\sigma^-_{l_4} + h.c.\nonumber\\
    &&-V \sum_{\square} P_{\sigma_{l_1}^x} P_{\sigma^x_{l_3}} +P_{\sigma_{l_2}^x}P_{\sigma^x_{l_4}}\nonumber\\
    && - \Gamma \sum_{s} e^{-i\alpha\left[1-\sum_{l\in s } P_{\sigma_l^x}\right]} + h.c. \label{eq:sqRK}
\end{eqnarray}
where we have defined the projection operator $P_\mathcal{O} = (1+\mathcal{O})/2$ for an operator $\mathcal{O}$ with eigenvalues $\pm 1$, $\Gamma=\infty$ enforces the hard-core constraint, and $\alpha$ can be any number (except for some special choices, such as $\pi$, for example).
The first sum is over square plaquettes on the lattice, and $l_{1\dots 4}$ are the four links going around clockwise or counterclockwise around it, and the second sum is over all sites which touch four links in a cross.

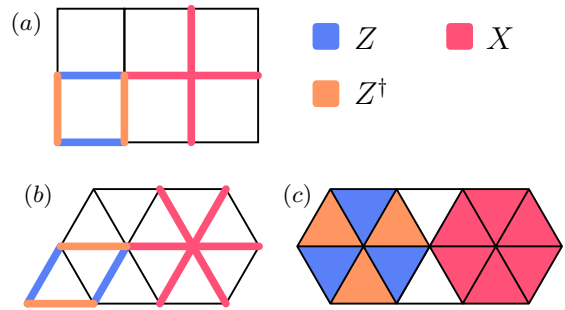
\begin{figure}
    \centering
    \begin{picture}(140,130)
    \put(-40,0){\includegraphics[width=0.4\textwidth]{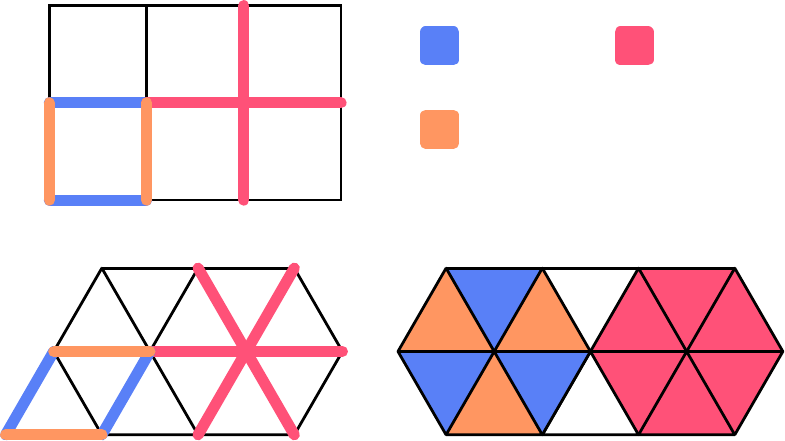}}
    \put(85,98){\large$Z$}
    \put(85,77.0){\large$Z^\dagger$}
    \put(135,98){\large$X$}
    \put(-45,105){$(a)$}
    \put(-40,40){$(b)$}
    \put(58,40){$(c)$}
    \end{picture}
    \caption{Pictorial representation of the terms in the Hamiltonian for $(a)$ the square lattice $N$-GDM, $(b)$ the triangular lattice $N$-GDM, and $(c)$ the triangular lattice $N$-GPM.
    Blue and orange bonds/triangles indicate operators involved in the kinetic terms in the Hamiltonian ($Z$ and $Z^\dagger$), and red indicates those involved in the site-constraint ($X$).
    Only one of three possible rhombus orientations is shown for the kinetic term in the triangular lattice $N$-GDM $(b)$.
    }
    \label{fig:ngdms}
\end{figure}

To arrive at the $N$-GDM on the square lattice, we first enlarge the $\mathbb{Z}_2$ degree of freedom on each bond to a $\mathbb{Z}_N$ degree of freedom, with operators $X$,$Z$ acting on them with algebra given in Eq~\ref{eq:znalg}.
We can then substitute $\sigma^+\rightarrow Z, \sigma^-\rightarrow Z^\dagger$ in the kinetic term of the RK Hamiltonian~\ref{eq:sqRK}.
Since the kinetic term does not annihilate any state, again the potential term is not needed.
We then have (schematically) 
\begin{equation}
    \mathcal{H}^{\text{Square}}_{N\text{-GDM}} = -\sum_{\square} (Z Z^\dagger Z Z^\dagger + h.c.)
    - \sum_{+} (\prod_{l\in +}X_{l} + h.c.)\label{eq:sqngdm}
\end{equation}
where we have suppressed the $l$ subscripts on the kinetic term which act on the four bonds around a square as illustrated in Figure~\ref{fig:ngdms}a.
The second term is the site constraint, which is a product over all four bonds emanating from a site.
This Hamiltonian is composed to mutually commuting terms (so we have set $t=\Gamma=1$) and can be solved exactly.
On the square lattice, this model is a $\mathbb{Z}_N$ generalization of the toric code~\cite{schulz}, which exhibits $\mathbb{Z}_N$ topological order as we will show.

For plaquette models, there is an additional difference between the $N$-GPM and the (hard-core) QPM in which kinetic terms are allowed.
In the QPM, the allowed flips $\mathcal{C}=\mathcal{C}_A\cup\mathcal{C}_B$ may only have each site being included in zero or two plaquettes, one from $\mathcal{C}_A$ and one from $\mathcal{C}_B$. 
In the $N$-GPM, the constraint is instead that each site only be a part of an equal number of triangles from $\mathcal{C}_A$ and $\mathcal{C}_B$.  
Thus, there are terms involving configurations where a site is included in more than two triangles total, that were not allowed in the QPM.

We shall now examine the properties of the $N$-GDM and $N$-GPM on a few characteristic lattices, starting with the square lattice $N$-GDM we just derived.

\subsection{$N$-GDM on Square Lattice}
On this (bipartite) lattice, the $N$-GDM is equivalent to a $\mathbb{Z}_N$ lattice gauge theory.
The Hamiltonian is given by Eq~\ref{eq:sqngdm}, and
we take the system on a torus which respects the bipartiteness of the square lattice.

The ground state degeneracy can be found by noting that for a non-contractible loop, the product $W = Z_{l_1}Z^\dagger_{l_2} \dots Z_{l_{L-1}} Z^\dagger_{l_L}$ along that loop commutes with and is independent of any of the terms in the Hamiltonian.
Furthermore, powers of $W$ are also independent of terms in the Hamiltonian. 
Since $W^{N}=1$, eigenstates may take on any eigenvalue $\omega^n$ for $n=1\dots N$.
As there are two such independent loop operators, the ground state sector is $N^2$-fold degenerate.

We can define the charge operator on site $s$ as 
\begin{equation}
    Q_s = 
    \begin{cases}
    \prod_{l\in s} X_l & s\in A\\
    \prod_{l\in s} X_l^{\dagger} & s \in B
    \end{cases}\label{eq:bipcharge}
\end{equation}
where $A$ and $B$ correspond to the two sublattices of the square lattice.
We then see that acting on the ground state with with $Z_l$ creates the exact eigenstate with two oppositely-charged excitations of charge $\omega$ and $\omega^{-1}$ on the two sites touching $l$.  
Therefore, total charge is preserved under any local operation modulo $N$.
Notice crucially that this construction works \emph{only} due to the bipartite nature of the lattice.  

Finally, we note that by doing a transformation $Z_l,X_l\rightarrow Z_l^{\dagger}, X_l^{\dagger}$ on a subset of the links, one can recover the usual form of the $\mathbb{Z}_N$ Toric code on the square lattice~\cite{schulz}.

\subsection{$N$-GDM on Triangular Lattice}
On non-bipartite lattices, the $N$-GDM describes a $\mathbb{Z}_2$ ordered phase for even $N$, and a topologically trivial liquid otherwise.
The Hamiltonian is 
\begin{equation}
    \mathcal{H}^{\text{Tri}}_{N\text{-GDM}} = -\sum_{\text{rhombus}}
    (Z Z^\dagger Z Z^\dagger + h.c.) - \sum_{s} (\prod_{l\in s} X_l + h.c.)
\end{equation}
where the first sum is now over length-4 loops on the triangular lattice which are rhombuses, and the second term is now a product over 6 links touching a site, which are illustrated in Figure~\ref{fig:ngdms}b.

We first consider the case of even $N$.
The first thing to note is that there is now an additional two-fold \emph{non}-topological ground state degeneracy.
We can write down the local operation $T_t = (Z_{l_1} Z_{l_2} Z_{l_3})^{N/2}$ where $l_{1\dots 3}$ are three links to go around a triangle $t$, which is independent of and commutes with the Hamiltonian.
Such triangle operators on different triangles can be related to each other via applications of terms in the Hamiltonian, and since $T_t^2=1$, there are degenerate ground states with $T_t=\pm 1$.
This is non-topological, as one can simply add a term $-h T_t$ to the Hamiltonian for just a single triangle, which would break the degeneracy.  We will ignore this degeneracy moving forwards.

Because the lattice is no longer bipartite, we cannot use the definition of charge from Eq.~\ref{eq:bipcharge}.
Instead, the best we can do is simply  
\begin{equation}
 Q_s = \prod_{l\in s} X_l.
 \label{eq:tringdmcharge}
\end{equation}
The action of applying $Z_l$ to a link $l$ creates two charges $\omega$ on each of the two sites it connects.  
As it is possible to locally create two charges $\omega^2$, a pair of such charges must be topologically indistinguishable from the vacuum.
In this case, we must make a distinction from the charge in Eq~\ref{eq:tringdmcharge} and the  \emph{topological} charge operator, which should be 
\begin{equation}
 Q^{\text{top}}_s = Q_s^{N/2}   ,
\end{equation}
 and can only take two values.
This is already an indication of the $\mathbb{Z}_2$ order to come, which we show by observing the $2^2$-fold topological degeneracy.

As before, consider the product  $W = Z_{l_1}Z^\dagger_{l_2} \dots Z_{l_{L-1}} Z^\dagger_{l_L}$ along a non-contractible loop of length $L$.
Again, $W$ is independent of and commutes with the Hamiltonian, so one might be tempted to say it can take on any of $N$ values.
However, this turns out not to be true, as $W^2$ \emph{can} be written as a product of terms in the Hamiltonian.
This is consistent with our previous finding that two charges are topologically identical to the vacuum: $W$ can be thought of as the process of moving a charge around the non-contractible loop, $W^2$ would correspond to moving two charges along the loop, which must therefore be trivial.
Since $W^2=1$ we are left with only a choice of $W=\pm 1$. 
There are two independent non-contractible loops, and so we are left with a $2^2$-fold topological degeneracy, for any even $N$.

For odd $N$, even a single charge must be topologically identical to the vacuum. 
To see this, observe that the local operator $(Z_{l_1} Z_{l_2}^{\dagger} Z_{l_3})^{(N+1)/2}$ for $l_{1\dots 3}$ going around a triangle creates a total charge $\omega$ on a single site, which therefore cannot carry any topological charge.

\subsection{$N$-GPM on Triangular Lattice}
We next consider $\mathbb{Z}_N$ generalized plaquette models ($N$-GPM).
Similar to how the properties of the $N$-GDM depended heavily on the bipartiteness of the lattice, we will find that the properties of the $N$-GPM with triangular plaquettes will depend heavily on the \emph{tri}partiteness of the lattice.

For this reason, we first examine the $N$-GPM on the triangular lattice, which has triangular plaquettes and is tripartite.
On this lattice, the $N$-GPM maps to a $\mathbb{Z}_N$ bosonic ring-exchange model on the (\emph{dual}) honeycomb lattice originally studied by Motrunich~\cite{motrunich} the strong coupling limit, which was found to have a fully deconfined $\mathbb{Z}_N\times \mathbb{Z}_N$ phase, which we will find here as well.

The Hamiltonian is
\begin{eqnarray}
    \mathcal{H}^\text{Tri}_{N\text{-GPM}} &=& -\sum_{s}(Z Z^\dagger Z Z^\dagger Z Z^\dagger + h.c.)\\ 
    && -\sum_{s}(X X X X X X + h.c.)\nonumber
\end{eqnarray}
where each term involves the product of operators over 6 triangles touching a site, as illustrated in Figure~\ref{fig:ngdms}c.
We again assume the system to be defined on a torus which respects the tripartiteness of the lattice.

Again, a simple method of analysis is by examining the quasiparticle structure.
Acting with $Z_t$ on a triangle creates three charge excitations, one on each sublattice which we label $A$, $B$, and $C$.
This leads to the ``fusion rule'' $a\times b\times c = 1$, where $a,b,c$ are charge excitations on each of the three sublattices.
Thus, we can represent $c$ as a bound state of an $a$ and $b$ antiparticle, and define the charge operators accordingly:
\begin{eqnarray}
    Q^{a}_s = 
    \begin{cases}
    \prod_{t\in s} X_t & s\in A\\
    \prod_{t\in s} 1 & s\in B\\
    \prod_{t\in s} X_t^{\dagger} & s \in C
    \end{cases}\nonumber
    \\
    Q^{b}_s = 
    \begin{cases}
    \prod_{t\in s} 1 & s\in A\\
    \prod_{t\in s} X_t & s\in B\\
    \prod_{t\in s} X_t^{\dagger} & s \in C
    \end{cases}\label{eq:tripcharge}
\end{eqnarray}
both of which are conserved under local operations.
Going through a similar exercise as before, one can readily verify the existence of \emph{four} independent non-contractible loop operators, which leads to the $N^2\times N^2$ topological ground state degeneracy.
These loop operators correspond to bringing an $a$ or $b$ particle around along a non-contractible loop.
For a more detailed analysis of this $\mathbb{Z}_N\times \mathbb{Z}_N$ phase, we direct the reader to Ref~\onlinecite{motrunich}, which discusses the model on the dual (honeycomb) lattice.

\begin{figure}
    \centering
    \begin{picture}(140,160)
    \put(-50,0){\includegraphics[width=0.5\textwidth]{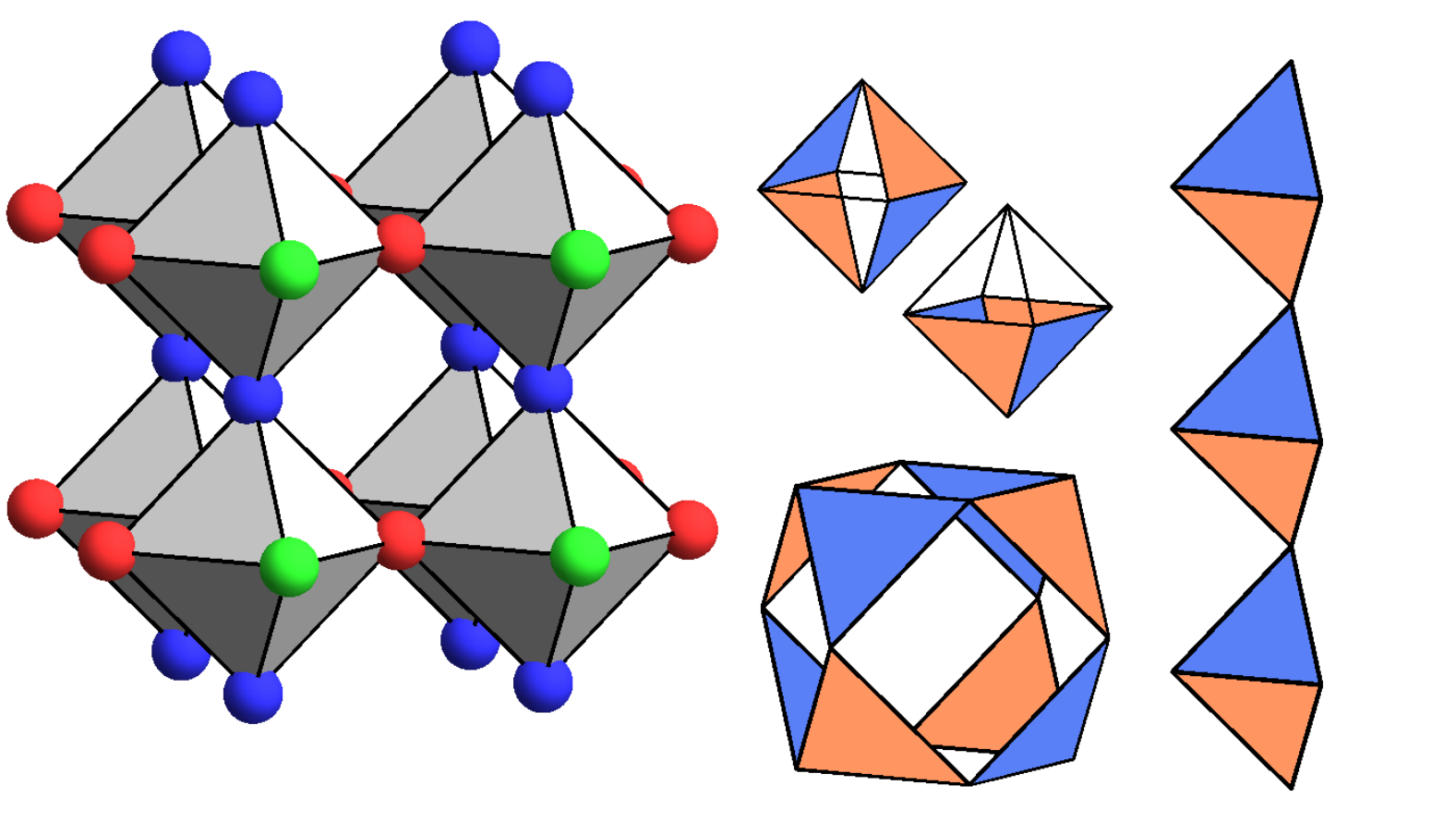}}
    \put(-50,130){$(a)$}
    \put(80,130){$(b)$}
    \put(80,70){$(c)$}
    \put(150,130){$(d)$}
    \put(168,118){$Z$}
    \put(168,102){$Z^\dagger$}
    \end{picture}
    \caption{
    The corner-sharing octahedra lattice, on which the $N$-GPM shows a $\mathbb{Z}_N$ X-cube fracton phase.  
    $(a)$ shows of the corner-sharing octahedra lattice, where sites from the three sublattices are colored red, green, and blue.
    The centers of the octahedra form into a simple cubic lattice, with lattice constant taken to be $1$.  
    Sites from each sublattice themselves also sit an offset simple cubic lattice.
    $(b)$ shows two type of   octahedron flips $|\mathcal{C}_\text{oct}|=4$, and $(c)$ shows a cuboctahedron flip $|\mathcal{C}_\text{cuboct}|=8$.  
    $(d)$ shows a portion of the $W_z(x_0,y_0)$ operator, which measures a $\mathbb{Z}_N$ topologically conserved quantity.
    Blue triangles indicates $Z$ operators and orange indicates $Z^\dagger$ operators.
    }
    \label{fig:cso}
\end{figure}
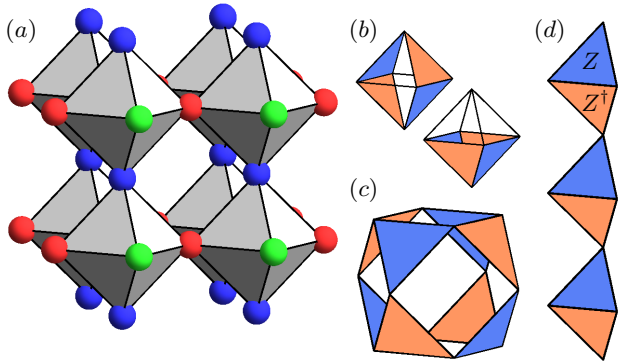
\subsection{$N$-GPM on Corner-Sharing Octahedra Lattice}
Here we highlight yet another interesting case: the $N$-GPM on the lattice defined by corner-sharing octahedra (a tripartite lattice with triangular plaquettes). 
The lattice can be understood as an underlying simple cubic lattice where each vertex is the center of an octahedron and the sites lie on the bonds of the underlying simple cubic lattice.
A portion of this lattice is shown in Figure~\ref{fig:cso}a, which also illustrates the tripartiteness of the lattice.
The $N$-GPM on this lattice will turn out to exhibit $\mathbb{Z}_N$ \emph{fracton} topological order, which appears to be described well by $\mathbb{Z}_N$ X-cube model~\cite{vijayhaahfu,vijay-layered}.
We will show that this model exhibits the key features of this phase: quasiparticle excitations which exhibit restricted movement and the characteristic subextensive topological ground state degeneracy.
Fundamental quasiparticle excitations of this (and the X-cube) model are the one-dimensionally mobile quasiparticle (which we call lineons~\cite{fractonfm}) and zero-dimensional immobile fractons, which are created at the corners of membrane operators.

The Hamiltonian describing this model is 
\begin{eqnarray}
\mathcal{H}^\text{C-S Oct}_{N\text{-GPM}} &=& -\sum_{\mathcal{C}_\text{oct}} \left(Z Z^\dagger Z Z^\dagger  + h.c.\right) \nonumber\\
&&- \sum_{\mathcal{C}_\text{cuboct}} \left( Z Z^\dagger Z Z^\dagger Z Z^\dagger Z Z^\dagger  + h.c.\right) \nonumber\\
&&-\sum_{s} \left(\prod_{t\in s} X_t + h.c. \right)
\end{eqnarray}
The first sum is over all bipartite sets of triangles $\mathcal{C}_\text{oct}=\mathcal{C}_A\cup\mathcal{C}_B$ that go around an octahedron, such that each site is a part of an equal number of triangles in $\mathcal{C}_A$ and $\mathcal{C}_B$, of size $|\mathcal{C}_\text{oct}|=4$  
These come in two main types, as shown in Figure~\ref{fig:cso}b (the rest are obtained by symmetry relations on the octahedron of these two).
The second sum is over all such sets on cuboctahedra (the 14-faced polyhedron with 8 triangular faces and 6 square faces), and involve all $|\mathcal{C}_\text{cuboct}|=8$ triangles, as shown in Figure~\ref{fig:cso}c.
Finally, the third term is the usual site constraint, with the product going over 8 triangles touching a site.

Again, we may begin our analysis by examining the quasiparticle structure.  
Apply a $Z_t$ to a triangle creates three charge excitations, one on each sublattice.
Let $A$,$B$, and $C$ correspond to the three sublattices, and $a$,$b$,and $c$ a single charge excitation on the respective sublattice.
We can apply the charge definition from Eq~\ref{eq:tripcharge} and treat the $c$ charge as a bound state of an $a$ and $b$ anticharge.
However, there is an additional conservation law here arising from the $\emph{geometry}$ of the lattice.  

Consider what happens when we have a single $a$ charge sitting on a site $s$ in the $A$ sublattice.
The simplest way it can be moved from $s$ to some other site $s^\prime$ is by applying the operator $Z^\dagger_{t_1}Z_{t_2}$, where $t_1$ must touch the site $s$ and share two sites with $t_2$, who must then touch another site $s^\prime$.
The geometry of the lattice allows only for $s^\prime$ to be one of two choices, which are both along one axis. 
Thus, this $a$ charge is confined to move along only one axis: it is the one-dimensional lineon of the X-cube model!
The $a$, $b$, and $c$ charges then correspond to lineons confined to move along $x$, $y$, and $z$ directions respectively.

The vison excitations can come in two forms: either as violations of the octahedron terms or as violations of the cuboctahedron terms.  
We first examine excitations of the cuboctahedron term:  
consider the operator $X_{t_1}X_{t_2}X_{t_3}X_{t_4}$ around the four triangles around a square-based pyramid (which comprises half of an octahedron).  
This operator commutes with every octahedron term, but creates four cuboctahedron excitations. 
Thus, cuboctahedron excitations can only be created in groups of four, and one can confirm that by repeated applications of this operator along a membrane, these excitations can be moved further apart and appear at the \emph{corners} of the membrane operator.
Alone, one such excitation cannot be moved without creating additional excitations.
The cuboctahedron vison excitations are therefore fractons!
Various combinations of octahedron excitations can then be interpreted as bound states of fracton excitations.

Finally, we can compute the ground state degeneracy.  
Consider the operator that corresponds to creating a $z$-moving lineon-antilineon pair at coordinates $(x_0,y_0)$, moving the lineon around in the positive $z$ direction, and then annihilating them again.  
This is done by a $Z Z^\dagger$ chain as shown in Figure~\ref{fig:cso}d, which we call $W_z(x_0,y_0)$ and commutes with the Hamiltonian.
Note that the details of how the $z$-lineon goes along each octahedron can be related to each other by octohedron terms in the Hamiltonian, and so are not independent.
We can henceforth freely choose $W_z(x_0,y_0)=\omega^n$ for $n=1\dots N$.
Furthermore, by application of the cuboctohedron term, we can show that in the ground state
\begin{equation}
     W_z(x_0,y_0)W_z^\dagger(x_0+1,y_0)W_z^\dagger(x_0,y_0+1)W_z(x_0+1,y_0+1) = 1
\end{equation}
where we have taken the length of the cubic unit cell to be 1, and so not all of these $W_z(x,y)$ are independent.
In fact, there are $2L-1$ independent $W_z(x,y)$'s, where $L$ is the linear dimension of the system.
To see this, let us define for convenience 
\begin{equation}
\tilde{W}_z(x,y) = 
    \begin{cases}
    W_z(x,y) & \text{if } x+y \text{ even}\\
    W_z^\dagger(x,y) & \text{if } x+y \text{ odd}
    \end{cases}
\end{equation}
Then, we can specify $2L-1$ of $\tilde{W}_z(x,y_0)$ and $\tilde{W}_z(x_0,y)$, and then obtain the rest via the relation 
\begin{equation}
    \tilde{W}_z(x,y) = \tilde{W}_z^\dagger(x,y_0)\tilde{W}_z^\dagger(x_0,y)\tilde{W}_z^\dagger(x_0,y_0).
\end{equation}
Therefore, we have $2L-1$ independent choices to make for the $z$ direction, and similarly along $x$ and $y$. 
This leads to a topological ground state degeneracy of $N^{6L-3}$, which for $N=2$ exactly matches with that of the X-cube model~\cite{vijayhaahfu}, despite being microscopically very different.
Thus, the $N$-GPM on the corner-sharing octahedra lattice results in $\mathbb{Z}_N$ fracton topological order, which appears to describe the same phase as the X-cube model.

\subsection{$N$-GPM on Simple Cubic Lattice}
Here, we briefly show how the $N$-GPM on the simple cubic lattice maps on to the $\mathbb{Z}_N$ X-cube model.
First, notice that this model has square plaquettes (thus describes a square tetramer model, rather than a trimer model).
The Hamiltonian is given by
\begin{equation}
    \mathcal{H}^{SC}_{N\text{-GPM}} = -\sum_{\text{matchboxes}}(Z Z^\dagger Z Z^\dagger + h.c.) - \sum_{s}\prod_{p\in s} X_p
\end{equation}
where the first sum is over four plaquettes going around a cube, which we refer to as ``matchboxes''.
There are three distinct orientations per cube.
To map the model on to the X-cube model, we transform to the dual lattice: cubic volumes are replaced by vertices, and plaquette faces are replaced by bonds.
The first sum then becomes the cross-term, and the second sum becomes the cube term.  
Finally, after mapping $Z\rightarrow Z^\dagger$ and $X\rightarrow X^\dagger$ for all operators on bonds going from A to B sublattices of the dual cubic lattice in the positive $\hat{x}$,$\hat{y}$, and $\hat{z}$ directions, one obtains the $\mathbb{Z}_N$ X-cube generalization obtained in Ref~\onlinecite{vijay-layered} from a layered construction.


\begin{thebibliography}{11}
\bibitem{qslreview} L. Savary and L. Balents Rep. Prog. Phys. 80 016502 (2016).
\bibitem{wen}   X.-G. Wen, Advances in Physics 44, 405 (1995).
\bibitem{haldane} F. D. M. Haldane and E. H. Rezayi, Phys. Rev. B 31 2529(R) (1985).
\bibitem{wen2}  X.-G. Wen and Q. Niu, Phys. Rev. B 41, 9377 (1990).
\bibitem{arovas} D. Arovas, J. R. Schrieffer, and F. Wilczek, Phys. Rev. Lett. 53, 722 (1984).
\bibitem{halperin} B.I. Halperin, Phys. Rev. B 25, (1982) 2185.
\bibitem{hamma} A. Hamma, R. Ionicioiu, and P. Zanardi,  Phys. Lett. A 337, 22 (2005).
\bibitem{kitaev} A. Kitaev and J. Preskill, Phys. Rev. Lett. 96, 110404 (2006).
\bibitem{levin} M. Levin and X-G. Wen, Phys. Rev. Lett. 96, 110405 (2006).
\bibitem{anderson2} P. Fazekas and P. W. Anderson, Philos. Mag. 30, 23 (1974).
\bibitem{anderson0}  P. W. Anderson 1973 Mater. Res. Bull. 8 153–60.
\bibitem{anderson}  P. W. Anderson, Science 235, 1196 (1987).
\bibitem{krs}  S. A. Kivelson, D. S. Rokhsar and J. P. Sethna, Phys. Rev. B 35, 8865 (1987).
\bibitem{qdmreview} R. Moessner and K. S. Raman, \emph{Introduction to Frustrated Magnetism (Berlin: Springer)} pp 437–79 (2011).
\bibitem{rk} D. S. Rokhsar and S. A. Kivelson, Phys. Rev. Lett. 61, 2376 (1998).
\bibitem{fradkin} E. Fradkin and S. A. Kivelson, Mod. Phys. Lett. B 4, 225 (1990).
\bibitem{igt} F. Wegner, J. Math. Phys. 12, 2259 (1971); R. Balian, J. M. Drouffe and C. Itzykson, Phys. Rev. D 11, 2098 (1975).
\bibitem{plaquettephase}  S. Sachdev, Phys. Rev. B 40, 5204 (1989); P. W. Leung, K. C. Chiu, and K. J. Runge, Phys. Rev. B 54, 12938 (1996); Olav F. Syljuasen, Phys. Rev. B 73, 245105 (2006).
\bibitem{cantordeconf} Eduardo Fradkin, David A. Huse, R. Moessner, V. Oganesyan, S. L. Sondhi,	Phys. Rev. B 69, 224415 (2004).
\bibitem{3dqdm} D. A. Huse, W. Krauth, R. Moessner, and S. L. Sondhi, Phys. Rev. Lett. 91, 167004 (2003).
\bibitem{3dqdm2} R. Moessner, and S. L. Sondhi, Phys. Rev. B 68, 184512 (2003).
\bibitem{3dqdm3} M. Hermele, M. P. A. Fisher, and L. Balents, Phys. Rev. B 69, 064404 (2004).
\bibitem{triangular-rvb} R. Moessner and S. L. Sondhi, Phys. Rev. Lett. 86, 1881 (2001).
\bibitem{qdpm} Owen Myers, C. M. Herdman, 	Phys. Rev. B 96, 174434 (2017).
\bibitem{pankov} S. Pankov, R. Moessner, and S. L. Sondhi Phys. Rev. B 76, 104436 (2007).
\bibitem{cenke} Cenke Xu and Congjun Wu, Phys. Rev. B 77, 134449 (2008).
\bibitem{vijayhaahfu} S. Vijay, J. Haah, and L. Fu, Phys. Rev. B 94, 235157 (2016).
\bibitem{williamson} D. J. Williamson, Phys. Rev. B 94, 155128 (2016).
\bibitem{fractonfm} T. Devakul, S. A. Parameswaran, S. L. Sondhi, 	arXiv:1709.10071.
\bibitem{fracton1} C. Chamon, Phys. Rev. Lett. 94, 040402 (2005).
\bibitem{fracton2} S. Bravyi, B. Leemhuis, B. Terhal, Ann. Phys. 326, 839 (2011).
\bibitem{fracton3} J. Haah, Phys. Rev. A 83, 042330 (2011).
\bibitem{fracton4} B. Yoshida, Phys. Rev. B 88, 125122 (2013).
\bibitem{fracton5} S. Vijay, J. Haah, and L. Fu, Phys. Rev. B 92, 235136 (2015).
\bibitem{rsf} R. Moessner and S. L. Sondhi and E. Fradkin, Phys. Rev. B. (2001).
\bibitem{senthilfisher} T. Senthil and M. P. A. Fisher,  Phys. Rev. B 62, 7850 (2000).
\bibitem{balents}  L. Balents, M. P. A. Fisher and C. Nayak, Int. J. Mod. Phys. B 12, 1033 (1998).
\bibitem{vison} T. Senthil and Matthew P. A. Fisher Phys. Rev. B 62, 7850 (2000).
\bibitem{motrunich} O. I. Motrunich, Phys. Rev. B 67, 115108 (2003).
\bibitem{motrunich2}  O. I. Motrunich, T. Senthil, Phys. Rev. Lett. 89, 277004 (2002)
\bibitem{kitaev-tc} A. Y. Kitaev,  2003, Ann. Phys. (N.Y.) 303, 2.
\bibitem{schulz} M. D. Schulz, S. Dusuel, R. Orus, J. Vidal, K. P. Schmidt, New J. Phys. 14, 025005 (2012)
\bibitem{vijay-layered} S. Vijay, arXiv:1701.00762.


\end{thebibliography}
\end{document}